# Improving adaptation of ubiquitous recommander systems by using reinforcement learning and collaborative filtering

Djallel Bouneffouf
Télécom SudParis
9, rue Charles Fourier
91011 Evry, France
Djallel.Bouneffouf@it-sudparis.eu

## ABSTRACT
The wide development of mobile applications provides a considerable amount of data of all types (images, texts, sounds, videos, etc.). Thus, two main issues have to be considered: assist users in finding information and reduce search and navigation time. In this sense, context-based recommender systems (CBRS) propose the user the adequate information depending on her/his situation. Our work consists in applying machine learning techniques and reasoning process in order to bring a solution to some of the problems concerning the acceptance of recommender systems by users, namely avoiding the intervention of experts, reducing cold start problem, speeding learning process and adapting to the user's interest.
To achieve this goal, we propose a fundamental modification in terms of how we model the learning of the CBRS. Inspired by models of human reasoning developed in robotic, we combine reinforcement learning and case-based reasoning to define a contextual recommendation process based on different context dimensions (cognitive, social, temporal, geographic). This paper describes an ongoing work on the implementation of a CBRS based on a hybrid Q-learning (HyQL) algorithm which combines Q-learning, collaborative filtering and case-based reasoning techniques. It also presents preliminary results by comparing HyQL and the standard Q-Learning w.r.t. solving the cold start problem.

## Categories and Subject Descriptors
H.3.3 [**Information Search and Retrieval**]: *information filtering, Selection process, Relevance feedback.*

## General Terms
Algorithms, Human factors

## Keywords
Context-based recommender systems machine learning application reasoning,, user personalization

## 1. INTRODUCTION
The need for adapting information systems to the user context has been accentuated by the extensive development of mobile applications that provide a considerable amount of data of all types (images, texts, sounds, videos, etc.). It becomes thus crucial to help users by guiding them in their access to information.

Systems should be able to recommend information helping the user to fulfill her/his goal and, thus, to accept the system. The information given by the system depends on the user's situation, i.e. an instance of the context. User's actions and associated situations reflect the user's information access interests. We consider four dimensions in the context: cognitive, temporal, geographic and social.

When applying techniques to adapt a recommender system to the user, major difficulties for ensuring the user acceptance follow.

- **Avoiding the intervention of experts**: when no initial information is given about the user, some systems use the intervention of an expert. However, on one hand, experts are not completely sure of the user's interest and may define wrong ideas about him; on the other hand, an expert is not always available.
- **Cold start:** In the initial state, without any information about the user and without the intervention of an expert, the system's behavior should not be incoherent for the user to not refuse it quickly.
- **A slow learning process:** the system's learning process about the user's interests has to be fast enough to avoid bothering the user with incorrect recommendations.
- **The evolution of the user's interest**: the interest of the user may change with time. The system has to be continuously adapted to this dynamicity using the user's context information to provide the relevant recommendations because, if the system's behavior is incoherent, the user refuses it quickly.

To better understand these problems, the following scenarios exemplify the usage of a recommender system.

**Scenario 1.** "Knowing the high mobility of its employees and their dependencies to the information contained in their corporate databases, the Nomalys company has equipped all mobile phones with the "NS" application. This application allows them to adapt to the nomadic life by consulting the company's database from their mobile phones. Because of the diversity of jobs existing in the company, Nomalys decides to provide the application with a generic recommender system, which has to retrieve the relevant information without any initial knowledge about users interests.

Paul, John and Lauren are new employees of the company integrating different teams (marketing, commercial, and technique resp.).

Regarding their agendas, they have a meeting with clients in Paris at midday. When they arrive at their meetings, the system should recommend them the relevant information which would help them to better manage their meeting. The system recommends Paul the register of complaints, John the register of factures and Lauren the technical registers."

To do these recommendations without the need of an expert and avoiding the cold start, the recommender system inferred them from the actions of the user's team, assuming they have the same interests.

**Scenario 2.** "Still in the same company, during one month, the recommender system finds that Paul often opens the register of complaints two hours before his meeting and not at the meeting. Moreover, John always tries to find companies which are near and do the same work as the one he will visit the next day. Using this knowledge, one month later, the system is able to recommend the register of complaints to Paul two hours before his meeting; it also recommends John companies which are near and do the same work as the one he will visit the next day".

In this case, the recommender system learned the interests of John and Paul through their actions in different situations, thus being able to recommend adequate information in similar new situations.

The system's learning process has to be fast to adequately follow the evolution of the user's interest.

In summary, in those scenarios, the recommender system starts with a predefined set of actions, not defined by an expert, but by the user's social group (in the scenario we talk about job teams). This initial default behavior allows the system to be ready-to-use. Then, the system is progressively adapted to a particular user using a learning lifelong process. Thus, the system is, at first, only acceptable to the user, and will, as time passes, give more and more satisfying results.

In the remaining of this paper, Section 2 is dedicated to related work. Then, in Section 3, we describe the current ideas of our ongoing work, followed by results concerning the cold start challenge in Section 4. Finally, we conclude, giving directions for future work.

## 2. Related work

The trend today on recommendation systems is to recommend relevant information to users, using supervised machine learning techniques. In these approaches, the recommender system has to pass by two steps:
(1) The learning step, where examples are presented to the system which "learns" from these examples and gradually adjusts its parameters to the desired output.
(2) The exploitation step, where new examples are presented to the system and exploitation asks it for generalization [9].
These approaches suffer from the following drawbacks:
  (i)   need of initial information about the user's interests provided by an expert;
  (ii)  slow learning about the user's behavior facing different situations;
  (iii) difficulty in following the changes of the user's interests. Some works found in the literature address these problems, as explained in what follows.

- **Avoiding the intervention of experts**: in [8, 24] the authors use Reinforcement Learning (RL) because it does not need previous experiences to start work. However, a major difficulty when applying RL techniques to real world problems is their slow convergence.
- Reducing **cold start problem:** Collaborative Filtering (CF) is often used to consider demographic information about users for providing them more accurate predictions [6, 25, 26]. However these techniques do not follow the user's interest evolution.
- **Accelerating the learning process**: different techniques are proposed to accelerate RL by using heuristics [18, 19] or case based reasoning in the robotic domain [10, 20, 21, 22]. As far as we know, none of them have been tested with recommender systems.
- **Adapting to the user's interest evolution:** the authors on [7, 17, 27] propose to follow the user's interests with an exploration strategy on the Q-learning algorithm. However, they do not address the remaining problems.

## 3. Proposition

Each work cited above (Section 2) tries to solve only one of the recommender systems' problems. In our work, we propose to address all of them as follows:
- **Avoiding the intervention of experts**: we propose to use the Q-learning algorithm which does not need initial user's information.
- Reducing **cold start problem:** we give Q-learning algorithm the ability to explore the knowledge of other users belonging to the same social network by using CF.
- **Accelerating the learning process:** to accelerate the Q-learning process, we mix Q-learning with case-based reasoning techniques to allow the reuse of cases and faster satisfy the user.
- **Adapting to the user's interest evolution**: we propose to use CF with an exploration strategy to follow the user's interest evolution.

Our proposition is based on three main algorithms. The following sections describe briefly their principle (Section 3.1 to Section 3.3) before presenting the complete proposed algorithm (Section 3.4).

## 3.1 Reinforcement learning and Q-learning algorithms

RL is a computational approach to learning whereby an agent tries to maximize the total amount of rewards it receives while interacting with a complex, uncertain environment [13]. A learning agent is modeled as a Markov decision process defined by (S, A, R, iR, P). S, A and R are finite sets of states, actions and rewards resp.; $iR : S \times A \rightarrow R$ is the immediate reward function and $P : S \times A \times S \rightarrow [0, 1]$ is the stochastic Markovian transition function. The agent constructs an optimal Markovian policy

$\pi : S \rightarrow A$ that maximizes the expected sum of future discounted rewards over an infinite horizon. We define $Q_\pi(s, a)$, the value of taking action *a* in state *s* under a policy $\pi$. The Q-learning algorithm allows computing an approximation of $Q_*$, independently of the policy being followed, if R and P are known. The Q-learning updated rule is:

$$Q_\pi(s,a) = Q_\pi(s,a) + \alpha \left[ r + \gamma \max_{a' \in A} Q_\pi(s',a') - Q_\pi(s,a) \right], \quad (1)$$

where *s* is the current state; *a* is the action performed in *s*; *r* is the reward received; *s'* is the next state; $\gamma$ is the discount factor ($0 \leq \gamma < 1$); and $\alpha$ is the learning rate.

In the Q-Learning algorithm, for every state *s*, action $a = Q(s)$ is chosen according to the current policy. The choice of the action by the policy must ensure a balance between exploration and exploitation phase.

The exploitation phase consists of choosing the best action for the current state, thus exploiting the system's knowledge. The exploration phase consists of choosing an action other than the best one in order to test it, observe its consequences, and increase the knowledge of the system.

There are several strategies to make the balance between exploration and exploitation. Here, we focus on two of them: the **greedy** strategy chooses always the best action from the Q-table, i.e. uses only exploitation; the **ε-greedy** strategy adds some greedy exploration policy, choosing a random action at each step if the policy returns the greedy action (probability = ε) or a random action (probability = *1 - ε*).

## 3.2 Collaborative filtering

A CF recommender system works as follows. Given a set of transactions *D*, where each transaction *T* identified by *id* is of the form <*id*, *item*, *rating*>, a recommender model *M* is produced. Each item is represented by a categorical value, while the rating is a numerical value in a given scale (e.g. each item is a movie rated with 1 to 5 stars). Such a model *M* can produce a list of *top-N* recommended items, and corresponding predicted ratings, from a given set of known ratings [4]. In many situations, ratings are not explicit. For example, if we want to recommend Web pages to a Web site visitor, we can use the set of pages she/he has visited, assigning those pages an implicit rate of one, and zero to all the other pages.

In terms of CF, three major classes of algorithms exist: *Memory-based*, *Model-based* and *Hybrid-based* [1, 4].

Memory-based approaches identify the similarity between two users by comparing their ratings on a set of items and have suffered from two fundamental problems: sparsity and scalability. Alternatively, the model-based approaches have been proposed to improve these problems, but these approaches tend to limit the range of users.

In our work, we use the *Hybrid-based* approach which combines the advantages of these two kinds of approaches by joining the two methods. Firstly, we employ memory-based CF to fill the vacant ratings of the user-item matrix. Then, we use the item-based CF as model-based to form the nearest neighbors of each item.

## 3.3 Case Based Reasoning

Case based reasoning (CBR) [11, 12] uses knowledge of previous situations (cases) to solve new problems, by finding a similar past case and reusing it in the new problem situation.

According to [11], solving a problem with CBR involves "obtaining a problem description, measuring the similarity of the current problem to previous problems stored in a case base with their known solutions, retrieving one or more similar cases, and attempting to reuse the solution of the retrieved case(s), possibly after adapting it to consider differences in problem descriptions". Some works found in the literature use CBR to implement recommender systems [15, 16].

## 3.4 The hybrid Q-learning

We improve the performance of the Q-learning in the following points:

- **Reusing past cases information:** to accelerate the Q-learning algorithm, we propose to integrate CBR into each iteration. Before choosing the best action, the algorithm computes the similarity between the present case and each one in the case base and, if there is a case that can be reused, the algorithm retrieves and adapts it.

- **Using social groups:** to give the Q-Learning the ability to use information from other users sharing the same interests, we propose to extend the ε-greedy strategy replacing the random action by another one that is selected computing the similarity of user profiles applying the CF algorithm as indicated in equation 2 where **q** is a random value uniformly distributed over [0, 1] and **p** (0≤**p**≤1) is a parameter that defines the exploration/exploitation tradeoff: the larger is **p**, the smaller is the probability of executing a random exploratory action. **a** $_{social\ group}$ is an action chosen among those available in state **s** by applying the CF algorithm.

$$\pi(s) = \begin{cases} \mathrm{argmax}_a\ Q(s, a) & \text{if } q \leq p, \\ a_{social\ group} & \text{otherwise} \end{cases} \quad (2)$$

The following algorithm is the proposed hybrid Q-learning algorithm, called HyQL:

```
Initialize Q(s, a) arbitrarily.
 Repeat (for each episode):
   Initialize s.
   Repeat (for each step):
     Compute similarity and cost.
   If there is a case that can be reused:
     Retrieve and adapt if necessary.
   Select an action a using equation 2.
   Execute the action a, observe r(s, a), s'.
   Update the values of Q(s, a) according to equation 1.
       s ← s'.
   Until s is terminal.
 Until some stopping criterion is reached.
```

## 4. Global mechanism

We are currently implementing a context-based recommender system (CBRS) which uses the HyQL algorithm. Figure 1 summarizes the global mechanism of the recommender system.

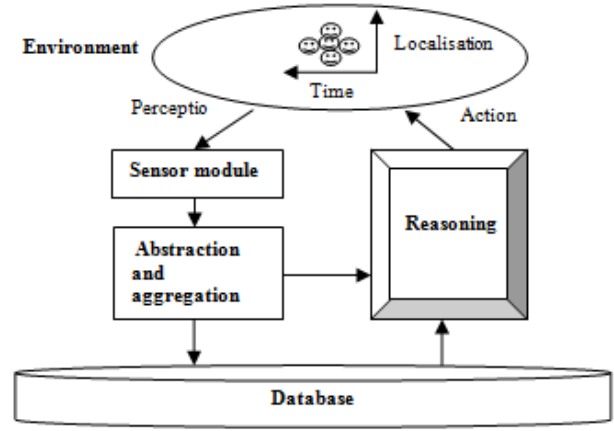

**Figure 1:** Global mechanism of the interaction between system and environment

The sensing module detects time, location, cognitive and social dimensions of the context in the following way:

(1) The cognitive dimension is given by all the actions of the user, like navigation (reads a document, opens a folder, etc.), sending an email and calling.

(2) The social group is predefined for each user. For example, all the marketing users of a company have the same need in general, thus they belong to the same group.

(3) Time is detected by the user's phone and the calendar of his/her institution.

(4) The geographic dimension is detected by the user's GPS.

In the thinking module, the abstraction phase is based on inference rules (e.g. specification / generalization) defined on the temporal and/or space ontologies. For instance, if we consider the outputs of GPS, we use an operation of "reverse geocoding" to get the corresponding address. The OWL-Time ontology [14] is today a reference for representing and reasoning about time. We propose to base our work on this ontology and extend it if necessary. Concerning the location ontology, we use a homemade one.

The aggregation phase is the combination of the two dimensions time and location, e.g. "morning at home." It describes situations in various granularity levels.

All modules of the system share a database divided into four parts:

*User:* describes and stores information about registered users.
*Preferences:* contains the couples (recommender system actions, user's rewards).
*Devices:* contains information about devices characteristics.
*History:* stores all occurred events and all actions taken by the system. This is useful for inferring the good recommendation to the user. It is divided into: **Action_history**, which contains all the interaction of the system with the environment; **Event_history**, which contains all the events registered by the user on his calendar.

The reasoning module chooses an action to deliver at each situation. In our experiments, the reasoning module is controlled by each of the previously presented algorithms: Q-Learning and HyQL.

## 5. Preliminary Results

At this stage of our work, we start by comparing Q-learning and HyQL w.r.t. solving the cold start problem. Our experiment is based on a simulation of the previously described scenarios. We consider two teams of ten users. Paul and John integrate each one a different team and are equipped with a Smartphone. We also consider that there are 100 resources in the user's DB.

To allow HyQL algorithm using CF, we gather John's team history of interactions with their phone.

The experiment consists of testing the first 100 trials of the system starting when the user is connected to the system. During the trials, the system has to recommend one resource from the database. We assume that, if the user chooses a recommended resource (e.g. reads a document, opens a folder, etc.), it is considered as a good recommendation.

To evaluate each trial, we use the traditional precision measure (percentage of good recommendations from all the recommendations done). Figure 3 shows the precision curves per 10 trial intervals for both Q-Learning and HyQL. These curves show that HyQL achieves better results than Q-Learning. In general, the precision of HyQL is greater than the precision of Q-Learning, except for trials number 50 where they have the same value.

This small experiment gives an indication about the better performance of HyQL comparing to Q-Learning in the case of reducing the cold start problem.

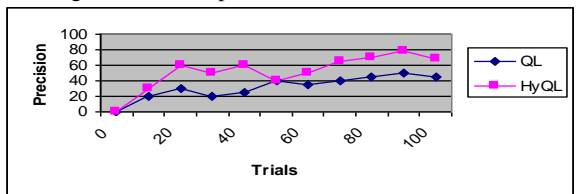

**Figure 3:** Percentage of good propositions in each trial using Q-learning and HyQL.

## 6. CONCLUSION

The aim of this work is to investigate the problems that we find when we try to adapt a recommender system to the user in a ubiquitous environment. The so-called context-based recommender system defines the observable situations and what actions should be executed in each situation in order to provide useful information to the user.

To achieve this goal, we propose to mix the RL algorithm with CBR and CF algorithms. The resulting hybrid algorithm (HyQL) gives good preliminary results compared with the standard Q-Learning. As future work, we intend to perform a full-simulated evaluation of HyQL w.r.t. other existing approaches, in a first step; then, to carry out tests with real users from Nomalys company.